\documentstyle[12pt]{article}

\newcommand{\kb}{\mbox{\boldmath $k$}}
\newcommand{\hA}{\hat{A}}
\newcommand{\hH}{\hat{H}}
\newcommand{\bbc}{\begin{center}}
\newcommand{\eec}{\end{center}}

\begin{document}

\begin{flushright}
Alberta Thy-09-97 \\
hep-ph/9705465 \\
August 1997\\
\end{flushright}

\vspace {0.3in}

\begin{center}
\LARGE Nonfactorization and Final State Interactions \\
\LARGE in $(B, B_s) \rightarrow \psi P$ and $\psi V\;$Decays\\
\vspace {0.3in}
{\large F. M. Al-Shamali and A. N. Kamal} \\ \vspace {0.1in}
 \small\em Theoretical Physics Institute and Department of Physics, \\
 \small\em University of Alberta, Edmonton, Alberta T6G 2J1, Canada.
\end{center}
\vspace {0.1in}

\begin{abstract}
Available experimental data on decay rates for  $B \rightarrow \psi K$, and decay rates and transversity amplitudes for  $B \rightarrow \psi K^*$ are used to investigate nonfactorization contributions in these decays using five different theoretical models for the formfactors. Using the knowledge on nonfactorization so gained, we study the processes $B_s \rightarrow \psi (\eta, \eta')$, $B_s \rightarrow \psi \phi$ and $B \rightarrow \psi(2S) K^*$. We find that present experimental data for the last two processes are consistent with the predictions of most of  the models considered.
\end{abstract}

\bbc (PACS numbers: 13.25.Hw, 14.40.Nd) \eec

\bbc \section{Introduction}
\eec
It was shown in \cite{ref:GKamalP94,ref:Aleksan95}  that factorization approximation used in conjunction with formfactors derived in most commonly used models failed to account for the ratio $(B \rightarrow \psi K)/(B \rightarrow \psi K^*)$ and the longitudinal polarization $\Gamma_L / \Gamma$ in $B \rightarrow \psi K^*$ decays. Subsequently, it was realized \cite{ref:Cheng94} that nonfactorized contributions could play an important role in these color-suppressed decays, and it was demonstrated \cite{ref:Cheng96,ref:KamalS96,ref:Cheng97} how such contributions could lead to an understanding of $B \rightarrow \psi K^*$ and $ \psi K$ data.

Our aim in this work is to investigate the role of nonfactorization and, where relevant, final state interactions {\em (fsi)}. The most recent CLEO data \cite{ref:CLEO97} enable us a complete amplitude analysis of $B \rightarrow \psi K^*$ decay. We use this to determine the three partial wave amplitudes, S, P and D, and the two relative phases. We exploit this knowledge in our work (see \cite{ref:Cheng97} for a more restricted analysis).

With the knowledge gained from the study of $B \rightarrow \psi K$ and $\psi K^*$ decays, we have also investigated the processes $B_s \rightarrow \psi \eta, \psi \eta', \psi \phi$ with special emphasis on the role of nonfactorization. Furthermore, we have extended our analysis to channels with $\psi(2S)$, instead of $\psi$, in the final state.

\bbc \section{$(B, B_s) \rightarrow \psi (\psi(2S)) P$ Decays}
\eec
We begin with the formulation for $B$ decays involving $\psi$ (or $\psi(2S)$) and a pseudoscalar particle in the final state. The decay amplitude, in the notation of  \cite{ref:KamalS96}, is written as
\begin{eqnarray}
A(B \rightarrow \psi P)  & = & \langle \psi P | {\cal H} ^{\mbox{eff}}_w
| B \rangle \nonumber\\
& = & \frac{G_F}{\sqrt{2}} V^*_{cb} V_{cs} 
\left[ a_2 \, \langle \psi P| (\overline{b} s) (\overline{c} c) | B \rangle
+ C_1 \, \langle \psi P|  {\cal H} ^{(8)}_w | B \rangle \right],
\label{eq:Amplitude}
\end{eqnarray}
where the brackets $(\overline{b} s)$ etc.\ represent $(V-A)$ quark currents and
\begin{eqnarray}
{\cal H} ^{(8)}_w & = & \frac{1}{2} \sum_a (\overline{b} \lambda^a s) (\overline{c} \lambda^a c),
\label{eq:Nonfac} \\
a_2 & = & \frac{C_1}{N_c} + C_2 . \label{eq:Wilsona2}
\end{eqnarray}
$N_c$ is the number of colors (taken to be 3) and $\lambda^a$ are the Gell-Mann matrices. $C_1$ and $C_2$ are the standard Wilson coefficients for which we take the values \cite{ref:KamalS96}, 
\begin{equation}
C_1 = 1.12 \pm 0.01, \hspace{0.75in} C_2 = -0.27 \pm 0.03,
\label{eq:WilsonC}\end{equation}
which are consistent with the choice in \cite{ref:Ruckl83}.

While the second term in (\ref{eq:Amplitude}), the matrix element of ${\cal H} ^{(8)}_w$, is nonfactorized, the first term receives both factorized and nonfactorized contributions \cite{ref:KamalSUV96}. We introduce the following definitions to proceed further,
\begin{eqnarray}
\langle \psi |  (\overline{c} c) | 0 \rangle & = & \epsilon^\mu m_\psi f_\psi ,
\label{eq:MED.a} \\
\langle P |  (\overline{b} s) | B \rangle & = & \left( p_B + p_P -
\frac{m^2_B - m^2_P}{q^2} q \right)_\mu F^{BP}_1(q^2) + \frac{m^2_B - m^2_P}{q^2} q_\mu F^{BP}_0(q^2) . \label{eq:MED.b} 
\end{eqnarray}
The first matrix element in (\ref{eq:Amplitude}) is then written as
\begin{eqnarray}
\langle \psi P | (\overline{b} s) (\overline{c} c) | B \rangle & = &
\langle P |  (\overline{b} s) | B\rangle \langle \psi |  (\overline{c} c) | 0 \rangle +
\langle P \psi | (\overline{b} s) (\overline{c} c) | B \rangle^{nf} \label{eq:MED.c} \\
 & \equiv & 2 m_\psi f_\psi C_P (\epsilon . p_B)  \left[F^{BP}_1(q^2) + F^{(1)nf}_1(q^2) \right], \label{eq:MED.d}
\end{eqnarray}
where we have defined
\begin{equation}
\langle \psi P | (\overline{b} s) (\overline{c} c) | B \rangle^{nf} =
2 m_\psi f_\psi C_P (\epsilon . p_B) F^{(1)nf}_1(q^2) . \label{eq:MED.e}
\end{equation}

In (\ref{eq:MED.d}) and (\ref{eq:MED.e}), $C_P$ has the following values,
\begin{equation}
C_P = \left\{
\begin{array}{cl}
\sqrt{\frac{2}{3}} \left( \cos\theta_P + \frac{1}{\sqrt{2}} \sin\theta_P
\right) & P \equiv \eta \\
\sqrt{\frac{2}{3}} \left( \frac{1}{\sqrt{2}} \cos\theta_P -  \sin\theta_P
\right) & P \equiv \eta' \\
1 & P \equiv K^0, K^+
\end{array} \right.
\end{equation}
with the $\eta - \eta'$ mixing angle $\theta_P = - 20^\circ$. The superscript (1) in
$F^{(1)nf}_1(q^2)$ denotes 'color-singlet'. The nonfactorized matrix element of
${\cal H} ^{(8)}_w$ is parametrized as
\begin{equation}
\langle \psi P| {\cal H} ^{(8)}_w | B \rangle^{nf} =
2 m_\psi f_\psi C_P (\epsilon . p_B) F^{(8)nf}_1(q^2) . \label{eq:MED.f}
\end{equation}

In terms of the definitions in (4) - (8), the decay amplitude for $B \rightarrow \psi P$ is written as
\begin{equation}
A(B \rightarrow \psi P) = \frac{G_F}{\sqrt{2}} V^*_{cb} V_{cs} a^{eff}_2  2 m_\psi f_\psi C_P
(\epsilon . p_B) F^{BP}_1(q^2) ,
\label{eq:MED.g}
\end{equation}
where
\begin{equation}
a^{eff}_2 = a_2 \left[ 1 + \frac{F^{(1)nf}_1(q^2)}{F^{BP}_1(q^2)} + \frac{C_1}{a_2}
\frac{F^{(8)nf}_1(q^2)}{F^{BP}_1(q^2)} \right].
\label{eq:MED.h}
\end{equation}

As a short-hand notation, we introduce two parameters $\chi_{F_1}$ and $\xi_{F_1}$ as measures of nonfactorized contributions, as follows:
\begin{equation}
a^{eff}_2 \equiv a_2 \left( 1 +  \frac{C_1}{a_2} \chi_{F_1} \right) \equiv a_2 \xi_{F_1} ,
\label{eq:a2eff}
\end{equation}
where,
\begin{equation}
\chi_{F_1} \equiv \frac{a_2}{C_1} \frac{F^{(1)nf}_1(q^2)}{F^{BP}_1(q^2)} + \frac{F^{(8)nf}_1(q^2)}{F^{BP}_1(q^2)} .
\end{equation}

Note that as $C_1/a_2$ is of the order of 10, the nonfactorized contribution from color-octet current is greatly enhanced. A departure of $\xi_{F_1}$ from unity, or $\chi_{F_1}$ from zero, signals nonfactorized contribution.

In terms of the quantities defined in (\ref{eq:MED.d}) - (\ref{eq:MED.h}), the decay rate for the exclusive channel $B \rightarrow \psi P$ is given by,
\begin{equation}
\Gamma (B \rightarrow \psi P) = \frac{G^2_F}{4 \pi} |V_{cb}|^2 |V_{cs}|^2
\left| a^{eff}_2 \right|^2 f^2_\psi  |C_P|^2 \nonumber |\kb|^3
\left| F^{BP}_1(m^2_\psi) \right|^2,
\label{eq:GammaBP}
\end{equation}
where $|\kb|$ is the momentum of the decay products in $B$ rest-frame.

Other parameters we used were, $V_{cs} = 0.974$, $V_{cb} = 0.04$ \cite{ref:Particle96}, $f_\psi = 0.384 \pm 0.014$ GeV and $f_{\psi(2S)} = 0.282 \pm 0.014$ GeV \cite{ref:Neubert92}.

\bbc \subsection{$B \rightarrow \psi(\psi(2S)) K$ decays}
\eec
We first consider the decay $B^+ \rightarrow \psi K^+$ whose branching ratio is more precisely measured than that of the neutral mode \cite{ref:Particle96}. The decay rate formula for this process, (\ref{eq:GammaBP}), can be rearranged and written as
\begin{equation}
\xi_{F_1} =  \frac{\sqrt{\Gamma (B^+ \rightarrow \psi K^+)}}{
(78.422 \times 10^{12} \mbox{ GeV}^{-2} \mbox{ sec}^{-1})^{1/2} \; |V_{cs}|
|V_{cb}| a_2 f_\psi F^{BK}_1(m_\psi^2)}  .
\label{eq:xiF1}
\end{equation}

A departure of $\xi_{F_1}$ from unity  signals the failure of the factorization assumption for a particular model-value of the formfactor. We employed five different models for $F^{BK}_1(m_\psi^2)$. They were: $(i)$ BSW I \cite{ref:BSW87-85}, where the formfactors are calculated at $q^2 = 0$ and extrapolated using a monopole form with the pole masses given in \cite{ref:BSW87-85}, $(ii)$ BSW II, where a dipole extrapolation is used for $F_1(q^2)$, $A_2(q^2)$ and $V(q^2)$, with the same pole masses as in \cite{ref:BSW87-85}, $(iii)$ CDDFGN \cite{ref:Casalbuoni93} , where the normalization of the formfactors are extrapolated using a monopole form, $(iv)$ AW \cite{ref:Altomari88}, where the formfactors are evaluated at the zero-recoil point corresponding to the maximum momentum transfer and then extrapolated down to the required $q^2$ using a monopole form, and $(v)$ ISGW \cite{ref:ISGW89}, where the formfactors are calculated at the maximum $q^2$ and extrapolated down to the needed value of $q^2$ with an exponential form. The predicted formfactors in these five models relevant to the processes of interest are shown in Tables \ref{tab:FormfactorF1} and \ref{tab:FormfactorA1}.

\begin{table}
\centering
\caption{Model predictions of formfactor $F_1(q^2)$ at $q^2 = m_\psi^2$ or
$m_{\psi(2S)}^2$. In CDDFGN model, $\eta$ stands for $\eta_8$, the octet
member. This scheme cannot handle $\eta_1$, the flavor singlet.}

\begin{tabular}{|l|ccccc|} \hline
& BSW I & BSW II & CDDFGN & AW & ISGW \\ \hline
$B^+ \rightarrow \psi K^+$ & 0.565 & 0.837 & 0.726 & 0.542 & 0.548 \\
$B^+ \rightarrow \psi(2S) K^+$ & 0.707 & 1.31 & 0.909 & 0.678 & 0.760 \\
$B^0_s \rightarrow \psi \eta$ & 0.49 & 0.726 & 0.771 & 0.534 & 0.293 \\
$B^0_s \rightarrow \psi(2S) \eta$ & 0.613 & 1.14 & 0.964 & 0.668 & 0.475 \\
$B^0_s \rightarrow \psi \eta'$ & 0.411 & 0.609 & --- & 1.06 & 0.463 \\
$B^0_s \rightarrow \psi(2S) \eta'$ & 0.514 & 0.954 & --- & 1.33 & 0.752 \\
\hline
\end{tabular}

\label{tab:FormfactorF1}
\end{table}
\begin{table}
\centering
\caption{Model predictions of $A_1(m_\psi^2), A_2(m_\psi^2),$ and $V(m_\psi^2)$ formfactors for the processes $B \rightarrow \psi K^*$,  $B \rightarrow \psi(2S) K^*$ and $B_s \rightarrow \psi \phi$}

\begin{tabular}{|l|l|ccc|cc|} \hline
\multicolumn{2}{|c|}{} & $A_1$ & $A_2$ & $V$ & $x$ & $y$ \\ \hline
& BSW I &0.458 & 0.462 & 0.548 & 1.01 & 1.19 \\
& BSW II & 0.458 & 0.645 & 0.812 & 1.41 & 1.77 \\
$B \rightarrow \psi K^*$ & CDDFGN & 0.279 & 0.279 & 0.904 & 1.00 & 3.24 \\
& AW & 0.425 & 0.766 & 1.19 & 1.80 & 2.81 \\
& ISGW & 0.316 & 0.631 & 0.807 & 2.00 & 2.56 \\ \hline
& BSW I &0.549 & 0.554 & 0.685 & 1.01 & 1.25 \\
& BSW II & 0.549 & 0.924 & 1.27 & 1.68 & 2.32 \\
$B \rightarrow \psi(2S) K^*$ & CDDFGN & 0.334 & 0.334 & 1.13 & 1.00 & 3.39 \\
& AW & 0.509 & 0.916 & 1.49 & 1.80 & 2.94 \\
& ISGW & 0.438 & 0.875 & 1.12 & 2.00 & 2.56 \\ \hline
& BSW I &0.374 & 0.375 & 0.466 & 1.00 & 1.24 \\
& BSW II & 0.374 & 0.523 & 0.691 & 1.40 & 1.85 \\
$B_s \rightarrow \psi \phi$ & CDDFGN & 0.265 & 0.279 & 0.919 & 1.05 & 3.47 \\
& AW & 0.449 & 0.703 & 1.34 & 1.56 & 2.98 \\
& ISGW & 0.237 & 0.396 & 0.558 & 1.67 & 2.35 \\
\hline
\end{tabular}

\label{tab:FormfactorA1}
\end{table}

\begin{figure}

\let\picnaturalsize=N
\def\picsize{3.4in}
\def\picfilename{figureA.eps}
\ifx\nopictures Y\else{\ifx\epsfloaded Y\else\input epsf \fi
\let\epsfloaded=Y
\centerline{\ifx\picnaturalsize N\epsfxsize \picsize\fi \epsfbox{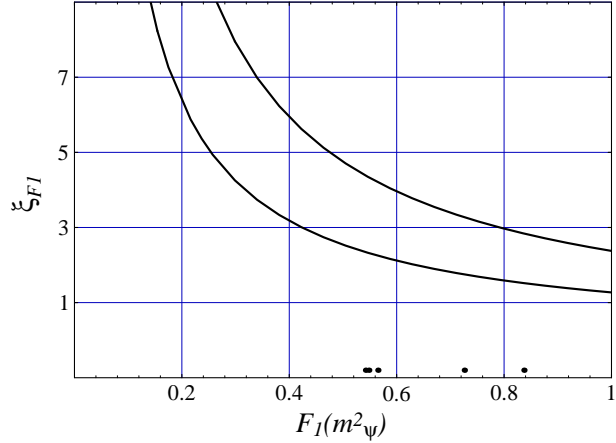}}}\fi

\caption{Allowed region (bounded by the two curves) of $\xi_{F_1}$ as a function of $F^{BP}_1(m_\psi^2)$ defined by $B^+ \rightarrow K^+ \psi$. The dots show the model predictions of the formfactors; from left to right: AW, ISGW, BSW I, CDDFGN, BSW II.}
\label{fig:A}
\end{figure}

We allowed $F^{BK}_1(m_\psi^2)$ to vary continuously and determined the allowed values of $\xi_{F_1}$ from data. The results are shown in Fig.\ \ref{fig:A}, where the dots represent the values of $F^{BK}_1(m_\psi^2)$ in various models which read from the left, AW, ISGW, BSW~I, CDDFGN, and BSW~II, in that order. $\xi_{F_1}$ different from unity (or $\chi_{F_1}$ different from zero), signals presence of nonfactorization contributions.

\begin{figure}

\let\picnaturalsize=N
\def\picsize{5.0in}
\def\picfilename{figureB.eps}
\ifx\nopictures Y\else{\ifx\epsfloaded Y\else\input epsf \fi
\let\epsfloaded=Y
\centerline{\ifx\picnaturalsize N\epsfxsize \picsize\fi \epsfbox{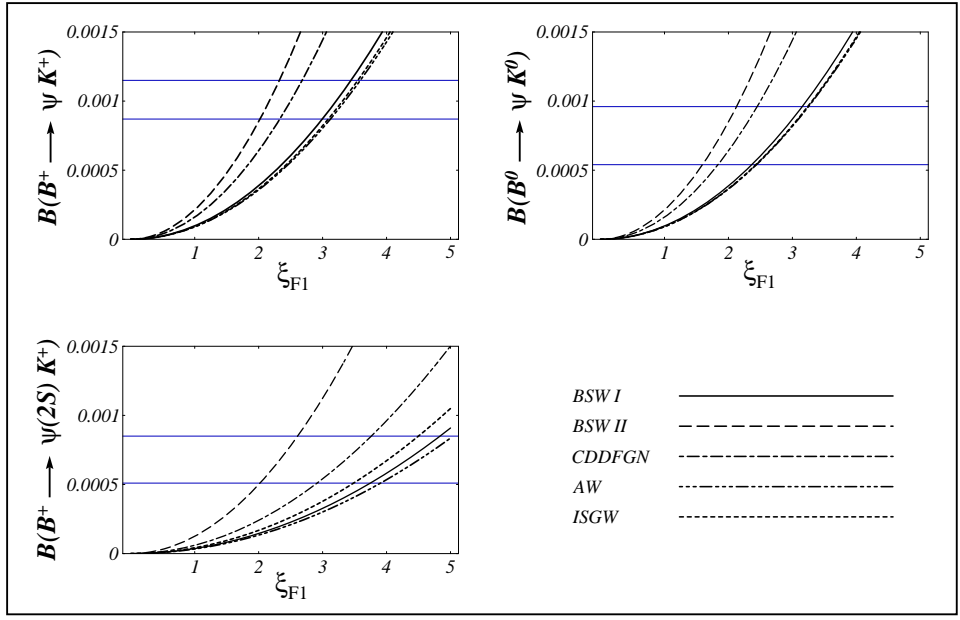}}}\fi

\caption{Branching ratios as functions of $\xi_{F_1}$ in each model. Horizontal lines define the branching ratio bounds to one standard deviation. Data from \protect\cite{ref:Particle96} for $B \rightarrow \psi K$ and \protect\cite{ref:CDF96-160E} for $B \rightarrow \psi(2S) K$}
\label{fig:B}
\end{figure}

We repeated the above analysis for $B^0 \rightarrow \psi \overline{K^0}$ and $B^{+} \rightarrow \psi(2S) K^{+}$. The results are displayed in the plots of Fig.\ \ref{fig:B}. For $B^{+} \rightarrow \psi K^{+}$, this is simply another way to display the results shown in Fig.\ \ref{fig:A}. In Fig.\ \ref{fig:B} we have plotted the branching ratios predicted in the five different models we have considered against the parameter $\xi_{F_1}$. One can read-off the amount of nonfactorization needed to understand the measured branching ratios in each model. Clearly, nonfactorized contributions are needed to explain data. For example, $B^{+} \rightarrow \psi  K^{+}$ branching ratio requires that the nonfactorization parameter $\xi_{F_1}$ is in the range (2 - 3.5), while $B^0 \rightarrow \psi  \overline{K^0}$ data require $\xi_{F_1}$ to be in the range (1.5 - 3). BSW II model requires the least amount of nonfactorization due to the fact that a dipole extrapolation of the formfactors allows the factorized term to be larger thereby reducing the necessity for the nonfactorized contribution.

\bbc \subsection{$B_s \rightarrow \psi (\psi(2S)) \eta, \eta'$ Decays}
\begin{figure}

\let\picnaturalsize=N
\def\picsize{5.0in}
\def\picfilename{figureC.eps}
\ifx\nopictures Y\else{\ifx\epsfloaded Y\else\input epsf \fi
\let\epsfloaded=Y
\centerline{\ifx\picnaturalsize N\epsfxsize \picsize\fi \epsfbox{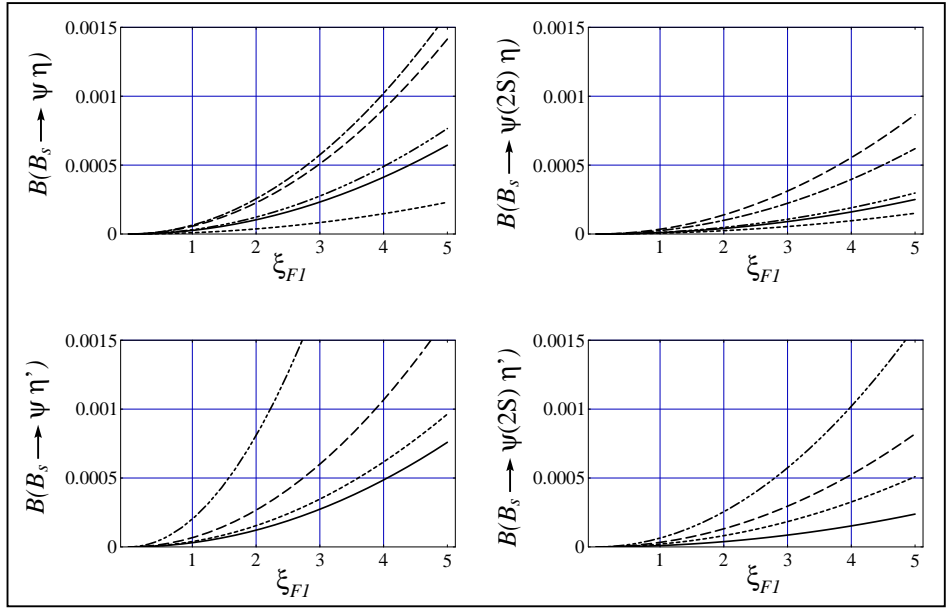}}}\fi

\caption{Branching ratios as a function of $\xi_{F_1}$ in each model.
In CDDFGN model, $\eta$ stands for $\eta_8$ and there is no prediction for $\eta'$.
See Fig.\ ~\protect\ref{fig:B} for legend.}
\label{fig:C}
\end{figure}

\eec
The analysis of the branching ratio data for $B \rightarrow \psi(\psi(2S) K$ decays can be used to predict the branching ratios for $B_s$ decays into $\psi$ or $\psi(2S)$ and $\eta$ or $\eta'$ if we assume that the amount of nonfactorized contribution is approximately independent of the light flavor. The calculation proceeds in a straight forward manner. We show the results in Fig. \ref{fig:C} where the branching ratios for $B_s \rightarrow (\psi \eta), (\psi(2S) \eta), (\psi \eta')$ or $(\psi(2S) \eta')$ are plotted as functions of $\xi_{F_1}$ for the five models we have considered.

For a given $\xi_{F_1}$, BSW II model produces the largest branching ratios for $B_s \rightarrow \psi(\psi(2S)) \eta$. For $B_s \rightarrow \psi(\psi(2S)) \eta'$ decays, the largest branching ratios for a given $\xi_{F_1}$ are generated in the AW model followed by those in BSW II scheme. In order to get some feel for the predicted branching ratios, we have presented the model-averaged branching ratios in Table \ref{tab:Bs-eta-psi} for $\xi_{F_1}$ = 1 (factorization),  $\xi_{F_1}$ = 2 and $\xi_{F_1}$ = 3.

\begin{table}[ht]
\centering
\caption{Average branching ratios predicted by the theoretical models for three
choices of $\xi_{F_1}$.}

\begin{tabular}{|l|c|c|c|} \hline
 & (Factorization) &  $(\xi_{F_1} = 2)$ &  $(\xi_{F_1} = 3)$ \\
 & $\times10^{-3}$ & $\times10^{-3}$ & $\times10^{-3}$ \\ \hline\hline
${\cal B}(B_s \rightarrow \psi \eta)$ &  0.037 & 0.15 & 0.33 \\
${\cal B}(B_s \rightarrow \psi(2S) \eta$) &  0.017 & 0.07 & 0.16 \\
${\cal B}(B_s \rightarrow \psi \eta'$) & 0.084 & 0.34 & 0.76 \\
${\cal B}(B_s \rightarrow \psi(2S) \eta'$) &  0.032 & 0.13 & 0.28 \\ \hline
\end{tabular}

\label{tab:Bs-eta-psi}
\end{table}

\bbc \section{$(B, B_s) \rightarrow \psi(\psi(2S) V$ Decays}
\eec
The decay amplitude for $B \rightarrow \psi V$ in the notation of \cite{ref:KamalS96} is
\begin{eqnarray}
A (B \rightarrow \psi V) & = & \frac{G_F}{\sqrt{2}} V_{cb} V_{cs}^* a_2 m_\psi f_\psi
\label{eq:AmplitudeBV} \\
& & \times \left\{ (m_B + m_V) (\epsilon^*_1 . \epsilon^*_2) \left(A^{BV}_1(m^2_\psi) + A^{(1)nf}_1 + \frac{C_1}{a_2} A^{(8)nf}_1 \right) \right. \nonumber \\
& & -\frac{(\epsilon^*_2 . (p_B - p_V)) (\epsilon^*_1 . (p_B + p_V))}{(m_B + m_V)}
\left(A^{BV}_2 (m^2_\psi) + A^{(1)nf}_2+ \frac{C_1}{a_2} A^{(8)nf}_2 \right) \nonumber \\
& & +\left. \frac{2i}{(m_B + m_V)}  \epsilon_{\mu \nu \alpha \beta} \epsilon^{*\mu}_1 \epsilon^{*\nu}_2 p^\alpha_V p^\beta_B
\left(V^{BV}(m^2_\psi) + V^{(1)nf}+ \frac{C_1}{a_2} V^{(8)nf} \right) \right\} \nonumber .
\end{eqnarray}
In (\ref{eq:AmplitudeBV}), $\epsilon_1$ and $\epsilon_2$ are the $\psi$ and $V$ polarization vectors respectively. $A^{BV}_1$, $A^{BV}_2$ and $V^{BV}$ are the formfactors defined in
\cite{ref:BSW87-85} which contribute to the factorized part of the decay amplitude. $A^{(8)nf}_1$, $A^{(8)nf}_2$ and $V^{(8)nf}$ are the nonfactorized contributions arising from the color-octet current products $(\overline{c} c) (\overline{s} b)$,  the analogues of $F^{(8)nf}_1$ in (\ref{eq:MED.f}). And $A^{(1)nf}_1$, $A^{(1)nf}_2$ and $V^{(1)nf}$ are the analogues of $F^{(1)nf}_1$ in (\ref{eq:MED.d}).

The following definitions \cite{ref:KamalS96}, in analogy with Eq.\ (\ref{eq:a2eff}), facilitate shorter forms for the equations that follow.
\begin{eqnarray}
\xi_i & = & 1 + \frac{A^{(1)nf}_i}{A^{BV}_i(m^2_{\psi})} + \frac{C_1}{a_2} \frac{A^{(8)nf}_i}{A^{BV}_i(m^2_{\psi})} \; \equiv \; 1 + \frac{C_1}{a_2} \chi_i,
\hspace{1.0cm} (i = 1, 2)
\nonumber\\
\xi_V & = & 1 + \frac{V^{(1)nf}}{V^{BV}(m^2_{\psi})} + \frac{C_1}{a_2} \frac{V^{(8)nf}}{V^{BV}(m^2_{\psi})} \; \equiv \; 1 + \frac{C_1}{a_2} \chi_V .
\label{eq:xis}
\end{eqnarray}

$B \rightarrow \psi V$ decays can be discussed in three equivalent basis-amplitudes: Helicity basis $(H_0, H_{+}, H_{-})$, Trasversity basis $(A_0, A_{||}, A_\perp)$, and Partial-wave basis $(S, P, D)$. They are related through the following definitions \cite{ref:Dighe96,ref:Wefollow},
\begin{eqnarray}
H_0 & = & - \frac{1}{\sqrt{3}} S + \sqrt{\frac{2}{3}} D \nonumber\\
H_+ & = & \frac{1}{\sqrt{3}} S + \frac{1}{\sqrt{2}} P + \frac{1}{\sqrt{6}} D \label{eq:Helicity}\\
H_- & = & \frac{1}{\sqrt{3}} S - \frac{1}{\sqrt{2}} P + \frac{1}{\sqrt{6}} D , \nonumber
\end{eqnarray}

\begin{eqnarray}
A_0 & = & H_0 = - \frac{1}{\sqrt{3}} S + \sqrt{\frac{2}{3}} D \nonumber\\
A_{||} & = & \frac{1}{\sqrt{2}} (H_{+} + H_{-}) =  \sqrt{\frac{2}{3}} S +  \frac{1}{\sqrt{3}} D \label{eq:Transversity} \\
A_\perp & = & \frac{1}{\sqrt{2}} (H_{+} - H_{-}) =  P . \nonumber
\end{eqnarray}

All amplitudes in (\ref{eq:Helicity}) and (\ref{eq:Transversity}) are, in principle, complex, their phases defined by the following (total angular momentum $J = 0$, uniquely determines the spin angular momentum once the orbital angular momentum is specified):
\begin{eqnarray}
S & = & |S| e^{i \delta_S} \nonumber\\
P & = & |P| e^{i \delta_P} \label{eq:SPD.1}\\
D & = & |D| e^{i \delta_D} \nonumber .
\end{eqnarray}

The process of generating complex amplitudes in terms of $\delta_S$, $\delta_P$ and $\delta_D$ is as follows: Helicity amplitudes are evaluated directly from (\ref{eq:AmplitudeBV}) {\em before final-state interaction (fsi) phases are put in}. This allows us to determine the {\em real} partial wave amplitudes, {\em before fsi}, through
\begin{eqnarray}
S & = & \frac{1}{\sqrt{3}} (H_{+} + H_{-} - H_0) \; = \;
\sqrt{\frac{2}{3}} A_{||} - \frac{1}{\sqrt{3}} A_0 \nonumber\\
P & = & \frac{1}{\sqrt{2}} (H_{+} - H_{-}) \; = \; A_\perp \label{eq:SPD.2}\\
D & = & \frac{1}{\sqrt{6}} (H_{+} + H_{-} + 2 H_0) \; = \;
\frac{1}{\sqrt{3}} A_{||} + \sqrt{\frac{2}{3}} A_0 .\nonumber
\end{eqnarray}
Once the real $S$, $P$ and $D$ amplitudes are determined, {\em their phase are put in by hand} as in (\ref{eq:SPD.1}), and with these complex $S$, $P$ and $D$ wave amplitudes one can write down complex $(H_0, H_{+}, H_{-})$ from (\ref{eq:Helicity}) or complex $(A_0, A_{||}, A_\perp)$ from (\ref{eq:Transversity}).

With these definitions the following expression for the decay rate are obtained
\begin{eqnarray}
\Gamma (B \rightarrow \psi V)
& = & \frac{|\kb|}{8 \pi m_B^2} (|H_0|^2 + |H_{+}|^2 + |H_{-}|^2) \nonumber\\
& = & \frac{|\kb|}{8 \pi m_B^2} (|A_0|^2 + |A_{||}|^2 + |A_\perp|^2) \nonumber\\
& = & \frac{|\kb|}{8 \pi m_B^2} (|S|^2 + |P|^2 + |D|^2) ,\label{eq:Gamma2}
\end{eqnarray}
where ${\kb}$ is the momentum in $B$ rest-frame.

The longitudinal and transverse polarizations are defined as,
\begin{eqnarray}
P_L & = & \frac{\Gamma_L}{\Gamma} = \frac{|H_0|^2}{|H_0|^2 + |H_{+}|^2 + |H_{-}|^2} \label{eq:PL}\\
P_T & = & 1 - P_L .\label{eq:PT}
\end{eqnarray}
We note from (\ref{eq:Helicity}) that both $P_L$ and $P_T$ depend on the relative phase between the $S$ and $D$ waves, $(\delta_{SD} = \delta_S - \delta_D)$, through $\cos \delta_{SD}$ in a compensatory manner such that $P_L + P_T = 1$. Alternatively, one can define in term of transversity amplitudes \cite{ref:Dighe96,ref:CLEO97},
\begin{eqnarray}
P_{||} & = & \frac{\Gamma_{||}}{\Gamma} \; = \;
\frac{|A_{||}|^2}{|A_0|^2 + |A_{||}|^2 + |A_\perp|^2} \label{eq:P{||}}\\
P_\perp & = & \frac{\Gamma_\perp}{\Gamma} \; = \;
\frac{|A_\perp|^2}{|A_0|^2 + |A_{||}|^2 + |A_\perp|^2} \label{eq:Pperp}\\
P_0 & = & 1 - P_{||} -P_\perp .\label{eq:P0}
\end{eqnarray}

From (\ref{eq:Transversity}), one notes that while $P_{||}$ depends on $\delta_{SD}$, $P_\perp$ is independent of the strong phases, $\delta_S$, $\delta_P$ and $\delta_D$. Eq.\ (\ref{eq:Helicity}) also shows that the only way information on $P$ wave phase can be obtained is via $\Gamma_{+} \propto |H_{+}|^2$ or $\Gamma_{-} \propto |H_{-}|^2$. $\Gamma_T = \Gamma_{+} + \Gamma_{-}$ is independent of the P-wave phase. With these definitions, we now consider the specific case of $B \rightarrow \psi K^*$.

\bbc \subsection{$B \rightarrow \psi K^*$ Decays}
\eec
With the definitions introduced in the preceding section, the helicity amplitudes for $B \rightarrow \psi K^*$ decays are (before {\em fsi} phases are introduced),
\begin{eqnarray}
H_0 & = & - \frac{G_F}{\sqrt{2}} V_{cb} V_{cs}^* f_\psi m_\psi (m_B + m_{K^*}) a_2
A^{BK^*}_1(m^2_\psi) (a \xi_1 - b \xi_2 x) \label{eq:Amplitude.3.1}\\
H_{\pm} & = & - \frac{G_F}{\sqrt{2}} V_{cb} V_{cs}^* f_\psi m_\psi (m_B + m_{K^*}) a_2
A^{BK^*}_1(m^2_\psi) (\xi_1 \mp c \xi_V y) \label{eq:Amplitude.3.2}
\end{eqnarray}
where \cite{ref:GKamalP94},
\begin{eqnarray}
a & = & \frac{m_B^2 - m_\psi^2 - m_{K^*}^2}{2 m_\psi m_{K^*}}, \nonumber\\
b & = & \frac{2 |\kb|^2 m_B^2}{m_\psi m_{K^*} (m_B + m_{K^*})^2}, \nonumber\\
c & = & \frac{2 |\kb| m_B}{(m_B + m_{K^*})^2}, \label{eq:abcxy}\\
x & = & \frac{A^{BK^*}_2(m_\psi^2)}{A^{BK^*}_1(m_\psi^2)}, \nonumber\\
y & = & \frac{V^{BK^*}(m_\psi^2)}{A^{BK^*}_1(m_\psi^2)} . \nonumber
\end{eqnarray}
With these helicity amplitudes, we can write the transversity amplitudes or the partial wave amplitudes via (\ref{eq:Transversity}) and (\ref{eq:SPD.2}).

We start with an amplitude analysis using the latest CLEO data \cite{ref:CLEO97}. Working in the transversity basis, they \cite{ref:CLEO97} determined the following branching ratio, dimensionless amplitudes (denoted by $\hA_0$, $\hA_\perp$ and $\hA_{||}$) and their phases,
\begin{eqnarray}
{\cal B}(B \rightarrow \psi K^*) & = & (1.35 \pm 0.18) \times 10^{-3} \nonumber\\
|\hA_0|^2 = \frac{|A_0|^2}{\Gamma} & = & 0.52 \pm 0.08 \nonumber\\
|\hA_\perp|^2 = \frac{|A_\perp|^2}{\Gamma} & = & 0.16 \pm 0.09 \label{eq:CLEO}\\
|\hA_{||}|^2 = \frac{|A_{||}|^2}{\Gamma} & = & 1 - |\hA_\perp|^2 - |\hA_0|^2 \; =\; 0.32 \pm 0.12 \nonumber\\
\phi_{||} & = & 3.00 \pm 0.37 \nonumber\\
\phi_\perp & = & -0.11 \pm 0.46, \nonumber
\end{eqnarray}
where $\phi_{||}$ and $\phi_\perp$ are the phases of the amplitudes $A_{||}$ and $A_\perp$, respectively, with the choice $\phi_0 = 0$ \cite{ref:CLEO97}. The important feature is that the amplitudes are relatively real.

Using the relation between the helicity, transversity and partial wave phases, the CLEO analysis (\ref{eq:CLEO}) can then be restated in the following equivalent forms,
\begin{eqnarray}
\mbox{Helicity basis:} & & \nonumber\\
|\hH_0|^2 = \frac{|H_0|^2}{\Gamma} & = & 0.52 \pm 0.08 \nonumber\\
|\hH_{+}|^2 = \frac{|H_{+}|^2}{\Gamma} & = & 0.014 \pm 0.034 \label{eq:CLEOhelicity}\\
|\hH_{-}|^2 = \frac{|H_{-}|^2}{\Gamma} & = & 1 - |\hH_0|^2 - |\hH_{+}|^2 \; =\; 0.47 \pm 0.08 \nonumber\\
\phi_{+} & = & 2.92 \pm 1.70 \nonumber\\
\phi_{-} & = & 3.01 \pm 0.29, \nonumber
\end{eqnarray}

\begin{eqnarray}
\mbox{Partial-wave basis:} & & \nonumber\\
|\hat{S}|^2 = \frac{|S|^2}{\Gamma} & = & 0.77 \pm 0.12 \nonumber\\
|\hat{P}|^2 = \frac{|P|^2}{\Gamma} & = & 0.16 \pm 0.09 \label{eq:CLEOpartial-waves}\\
|\hat{D}|^2 = \frac{|D|^2}{\Gamma} & = & 1 - |\hat{S}|^2 - |\hat{P}|^2 \; =\; 0.073 \pm 0.044 \nonumber\\
\phi_S & = & 3.07 \pm 0.19 \nonumber\\
\phi_P & = & -0.11 \pm 0.46 \nonumber\\
\phi_D & = & 0.17 \pm 0.44, \nonumber
\end{eqnarray}
The phases in (\ref{eq:CLEOhelicity}) and (\ref{eq:CLEOpartial-waves}) are evaluated relative to $\phi_0$ as in the CLEO analysis \cite{ref:CLEO97}. From (\ref{eq:CLEOpartial-waves}) we note that $|S| > |P| > |D|$ as one might intuitively anticipate.

Next, having determined the helicity amplitudes, we are in a position to extract information on the parameters $\xi_1$, $\xi_2$ and $\xi_V$ in a given model for the formfactors. We summarize our method below.

The helicity amplitudes in terms of $\xi_1$, $\xi_2$ and $\xi_V$ are given in Eqs.\ (\ref{eq:Amplitude.3.1}) and (\ref{eq:Amplitude.3.2}). The parameters $\xi_1$, $\xi_2$ and $\xi_V$, representing nonfactorization, are then obtained from the constraints,
\begin{eqnarray}
|\hH_0|^2 & = & \frac{(a - b x \xi_{21})^2}{2 (1 + c^2 y^2 \xi^2_{V1}) + (a - b x \xi_{21})^2} =
0.52 \pm 0.08 \label{eq:H02}\\
|\hH_{+}|^2 & = & \frac{(1 - c y \xi_{V1})^2}{2 (1 + c^2 y^2 \xi^2_{V1}) + (a - b x \xi_{21})^2} =
0.014 \pm 0.034 \label{eq:Hplus2}
\end{eqnarray}
and
\begin{eqnarray}
{\cal B}(B \rightarrow \psi K^*) & = & \frac{G_F^2 |{\kb}|}{16 \pi m_B^2} |V_{cb}|^2 |V_{cs}|^2 f_\psi^2 m_\psi^2 (m_B + m_{K^*})^2 a^2_2 |A_1(m^2_\psi)|^2 \xi_1^2 \nonumber\\
& & \times \left\{ (a - b x \xi_{21})^2 + 2 (1 + c^2 y^2 \xi^2_{V1}) \right\} \nonumber\\
& = & (1.35 \pm 0.18) \times 10^{-3}
\label{eq:BR}
\end{eqnarray}
where we have defined the ratios,
\begin{equation}
\xi_{21} = \frac{\xi_2}{\xi_1} \hspace{1.0cm} \mbox{and} \hspace{1.0cm}
\xi_{V1} = \frac{\xi_V}{\xi_1}.
\end{equation}

\begin{figure}

\let\picnaturalsize=N
\def\picsize{3.4in}
\def\picfilename{figureD.eps}
\ifx\nopictures Y\else{\ifx\epsfloaded Y\else\input epsf \fi
\let\epsfloaded=Y
\centerline{\ifx\picnaturalsize N\epsfxsize \picsize\fi \epsfbox{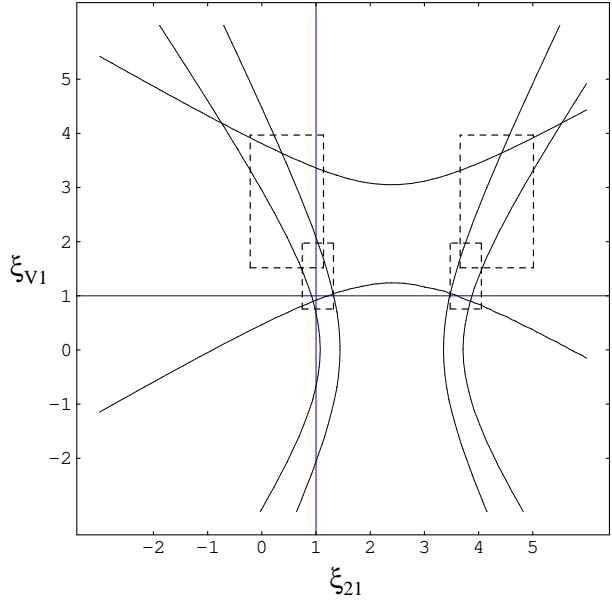}}}\fi

\caption{The region in $\xi_{21}$ and $\xi_{V1}$ plane allowed by the latest CLEO II measurments of $|\hH_0|^2$ (vertical bands) and $|\hH_{+}|^2$ (horizontal band) for $B \rightarrow \psi K^*$ in BSW I model.}
\label{fig:E}
\end{figure}

The regions in $\xi_{21}$ and $\xi_{V1}$ space that explain the experimental value of polarization $(|\hH_0|^2)$ are shown by the two vertical bands in Fig.\ \ref{fig:E}, while the region between the two horizontal curves corresponds to non-negative values for $|\hH_+|^2$ within the error. The boxes in Fig.\ \ref{fig:E}  show the four solutions we get by solving (\ref{eq:H02}) and (\ref{eq:Hplus2}) for $\xi_{21}$ and $\xi_{V1}$. The errors in $\xi_{21}$ and $\xi_{V1}$ are correlated as parts of the boxes lie ouside the overlap of the horizontal and vertical bands.

Clearly, in BSW~I model, within errors, there are solutions with $\xi_{21} = 1$ and $\xi_{V1} = 1$ i.e.\ $\xi_1 = \xi_2 = \xi_V$. This is the class of solutions discussed in \cite{ref:Cheng97}. The value of $\xi_1$ is then obtained from (\ref{eq:BR}) and does not allow $\xi_1 = 1$ solution. We repeated this procedure for other models of the formfactors. the results are tabulated in Table 4.

\begin{table}
\centering
\caption{Solutions of $\xi_1$, $\xi_{21}$ and $\xi_{V1}$ using the latest CLEO II measurements of branching ratio, polarization and $|\hH_{+}|^2$ for the process $B \rightarrow \psi K^*$. In the table only positive solutions for $\xi_1$ are shown since for every solution of $\xi_1$ there is another which is its negative.}

\begin{tabular}{|l|l|l|l|l|l|} \hline
\multicolumn{2}{|l|}{Model} & Solution 1 & Solution 2 & Solution 3 & Solution 4 \\ \hline\hline
& $\xi_1 $& $1.79 \pm 0.62$ & $1.79 \pm 0.62$ & $1.26 \pm 0.55$ & $1.26 \pm 0.55$ \\
BSW I &  $\xi_{21}$ & $1.03 \pm 0.29$ & $3.77 \pm 0.29$ & $0.46 \pm 0.68$ &
$4.34 \pm 0.68$ \\
& $\xi_{V1}$ & $1.36 \pm 0.61$ & $1.36 \pm 0.61$ & $2.74 \pm 1.20$ & $2.74 \pm 1.20$ \\ \hline\hline
& $\xi_1 $ & $1.79 \pm 0.62$ & $1.79 \pm 0.62$ & $1.26 \pm 0.55$ & $1.26 \pm 0.55$ \\
BSW II &  $\xi_{21}$ & $0.74 \pm 0.21$ & $2.70 \pm 0.21$ & $0.33 \pm 0.49$ &
$3.11 \pm 0.49$ \\
& $\xi_{V1}$ & $0.92 \pm 0.41$ & $0.92 \pm 0.41$ & $1.84 \pm 0.83$ &
$1.84 \pm 0.83$ \\ \hline\hline
& $\xi_1 $ & $2.94 \pm 1.00$ & $2.94 \pm 1.00$ & $2.07 \pm 0.91$ & $2.07 \pm 0.91$ \\
CDDFGN &  $\xi_{21}$ & $1.04 \pm 0.29$ & $3.80 \pm 0.29$ & $0.46 \pm 0.68$ &
$4.38 \pm 0.68$ \\
& $\xi_{V1}$ & $0.50 \pm 0.22$ & $0.50 \pm 0.22$ & $1.01 \pm 0.45$ & $1.01 \pm 0.45$ \\ \hline\hline
& $\xi_1 $ & $1.93 \pm 0.67$ & $1.93 \pm 0.67$ & $1.36 \pm 0.60$ & $1.36 \pm 0.60$ \\
AW &  $\xi_{21}$ & $0.58 \pm 0.16$ & $2.11 \pm 0.16$ & $0.26 \pm 0.38$ &
$2.43 \pm 0.38$ \\
& $\xi_{V1}$ & $0.58 \pm 0.26$ & $0.58 \pm 0.26$ & $1.16 \pm 0.52$ &
$1.16 \pm 0.52$ \\ \hline\hline
& $\xi_1 $ & $2.60 \pm 0.90$ & $2.60 \pm 0.90$ & $1.83 \pm 0.80$ & $1.83 \pm 0.80$ \\
ISGW &  $\xi_{21}$ & $0.52 \pm 0.15$ & $1.90 \pm 0.15$ & $0.23 \pm 0.34$ &
$2.19 \pm 0.34$ \\
& $\xi_{V1}$ & $0.63 \pm 0.28$ & $0.63 \pm 0.28$ & $1.27 \pm 0.57$ & $1.27 \pm 0.57$ \\
\hline
\end{tabular}

\label{tab:nofactorizatiosAV}
\end{table}

From Table 4 it is evident that only BSW I model permits solutions with $\xi_1 = \xi_2 = \xi_V$ (but $\xi_1 \neq 1$); other models do not allow solutions with $\xi_1 = \xi_2 = \xi_V$. There are  no solutions to the latest CLEO data \cite{ref:CLEO97} consistent with factorization i.e., $\xi_1 = \xi_2 = \xi_V = 1$ in any of the models we have considered.

We chose to work with CLEO data only rather than use the world average for the longitududinal polarization because it is only the CLEO data which allow a complete determination of the decay amplitude. For the record , the world average (our estimate) of all measurements \cite{ref:ARGUS94,ref:CLEO94,ref:CDF95,ref:CLEO97} of the longitududinal polarization in $B \rightarrow \psi K^*$ is
\begin{equation} P_L = 0.66 \pm 0.05 .\end{equation}
To one stadard deviation there is no overlap of the world average and the CLEO data. However, within errors there are values of $\xi_{21}$ and $\xi_{V1}$, shown in Table \ref{tab:nofactorizatiosAV} that fit the world-averaged $P_L$.

\bbc \subsection{$B_s \rightarrow \psi \phi$ Decays}
\eec
If it is assumed that that nonfactorization contributions are independent of the flavor of the light degree of freedom we can use the values of $\xi_1$, $\xi_{2,1}$ and $\xi_{V,1}$ in Table \ref{tab:nofactorizatiosAV} to predict the branching ratio, polarization and transversity for  $B_s \rightarrow \psi \phi$. Indeed, the branching ratio, longitudinal polarization and trasverse polarization $(|\hH_-|^2)$ for $B_s \rightarrow \psi \phi$ can be related directly to the those for $B \rightarrow \psi K^*$ by eliminating  $\xi_1$, $\xi_{21}$ and $\xi_{V1}$. The result is shown in Table~\ref{tab:B-psi-phi}. The following experimental data for $B_s \rightarrow \psi \phi$ are now available \cite{ref:CDF95,ref:CDF96}
\begin{eqnarray}
{\cal B}(B_s \rightarrow \psi \phi) & = & ( 0.93 \pm 0.33) \times 10^{-3} \nonumber\\
P_L = |\hH_0|^2 & = & 0.56 \pm 0.21 .
\end{eqnarray}
\begin{table}
\centering
\caption{The branching ratios, $|\hH_0|^2$ and $|\hH_-|^2$ for the process $B_s \rightarrow \psi \phi$ using for $\xi_1$, $\xi_{21}$ and $\xi_{V1}$ the values in
Table \protect\ref{tab:nofactorizatiosAV} calculated for $B \rightarrow \psi K^*$.}

\begin{tabular}{|l|l|l|l|l|l|} \hline
\multicolumn{2}{|l|}{Model} & Solution 1 & Solution 2 & Solution 3 & Solution 4 \\ \hline\hline
& BR $\times10^{-3}$&
$0.91 \pm 0.14$ & $0.75 \pm 0.12$ & $0.88 \pm 0.14$ & $0.77 \pm 0.12$ \\
BSW I &  $|\hH_0|^2$ &
$0.50 \pm 0.07$ & $0.39 \pm 0.08$ & $0.48 \pm 0.07$ & $0.40 \pm 0.08$ \\
& $|\hH_-|^2$ &
$0.49 \pm 0.08$ & $0.60 \pm 0.09$ & $0.50 \pm 0.08$ & $0.58 \pm 0.10$ \\ \hline\hline
& BR $\times10^{-3}$ &
$0.91 \pm 0.14$ & $0.75 \pm 0.12$ & $0.89 \pm 0.14$ & $0.78 \pm 0.12$ \\
BSW II &  $|\hH_0|^2$ &
$0.49 \pm 0.07$ & $0.39 \pm 0.08$ & $0.48 \pm 0.08$ & $0.41 \pm 0.08$ \\
& $|\hH_-|^2$ &
$0.49 \pm 0.08$ & $0.59 \pm 0.09$ & $0.51 \pm 0.08$ & $0.58 \pm 0.10$ \\ \hline\hline
& BR $\times10^{-3}$ &
$1.20 \pm 0.19 $ & $1.20 \pm 0.19 $ & $ 1.20 \pm 0.19 $ & $ 1.20 \pm 0.19 $ \\
CDDFGN &  $|\hH_0|^2$ &
$0.47 \pm 0.08$ & $0.47 \pm 0.08$ & $0.46 \pm 0.08$ & $0.46 \pm 0.08$ \\
& $|\hH_-|^2$ &
$0.52 \pm 0.09$ & $0.52 \pm 0.09$ & $0.52 \pm 0.09$ & $0.52 \pm 0.09$ \\ \hline\hline
& BR $\times10^{-3}$ &
$1.70 \pm 0.28$ & $0.96 \pm 0.17$ & $1.50 \pm 0.27$ & $1.00 \pm 0.18$ \\
AW &  $|\hH_0|^2$ &
$0.53 \pm 0.06$ & $ 0.19 \pm 0.08$ & $0.49 \pm 0.07$ & $0.24 \pm 0.09$ \\
& $|\hH_-|^2$ &
$0.45 \pm 0.08$ & $ 0.79 \pm 0.08$ & $0.50 \pm 0.06$ & $0.73 \pm 0.13$ \\ \hline\hline
& BR $\times10^{-3}$ &
$0.82 \pm 0.15$ & $0.42 \pm 0.07$ & $0.71 \pm 0.14$ & $0.43 \pm 0.07$ \\
ISGW &  $|\hH_0|^2$ &
$0.57 \pm 0.06$ & $0.15 \pm 0.08$ & $0.54 \pm 0.06$ & $0.23 \pm 0.11$ \\
& $|\hH_-|^2$ &
$0.41 \pm 0.07$ & $0.80 \pm 0.08$ & $0.46 \pm 0.06$ & $0.76 \pm 0.13$ \\ \hline\hline
\multicolumn{2}{|l|}{Experiment \cite{ref:CDF95,ref:CDF96}} &
\multicolumn{4}{|l|}{ ${\cal B}(B_s \rightarrow \psi \phi) = ( 0.93 \pm 0.33) \times 10^{-3}$} \\
\multicolumn{2}{|l|}{} & \multicolumn{4}{|l|}{ $P_L = |\hH_0|^2 = 0.56 \pm 0.21$} \\ \hline
\end{tabular}

\label{tab:B-psi-phi}
\end{table}

By studying the results presented in Table~\ref{tab:B-psi-phi}, we note the following: {\em (i)} Most of the predictions (those given by BSW I model, BSW II model, CDDFGN model and both first and third solutions of the ISGW model) are consistent with available experimental results. {\em (ii)} The predictions of the BSW I, BSW II and CDDFGN models show very little sensitivity to the solution-type. {\em (iii)}  $|\hH_+|^2$ has a very small value in all the formfactor models considered.

\bbc \subsection{$B \rightarrow \psi(2S) K^*$ Decays}
\eec
The branching ratio, longitudinal polarization and transverse polarization $(|\hH_-|^2)$ for $B \rightarrow \psi(2S) K^*$ can also be related directly to those for $B \rightarrow \psi K^*$ by eliminating $\xi_1$, $\xi_{21}$ and $\xi_{V1}$. The result is shown in
Table~\ref{tab:B-psi2S-Kstar}. To the best of our knowledge, only the branching ratio for this process is available \cite{ref:CDF96-160E},
\begin{equation}
{\cal B} (B\rightarrow \psi(2S) K^*) = ( 0.9 \pm 0.29) \times 10^{-3}.
\end{equation}
\begin{table}
\centering
\caption{The branching ratios, $|\hH_0|^2$ and $|\hH_-|^2$ for the process $B \rightarrow \psi(2S) K^*$ using for $\xi_1$, $\xi_{21}$ and $\xi_{V1}$ the values in
Table \protect\ref{tab:nofactorizatiosAV} calculated for $B \rightarrow \psi K^*$.}

\begin{tabular}{|l|l|l|l|l|l|} \hline
\multicolumn{2}{|l|}{Model} & Solution 1 & Solution 2 & Solution 3 & Solution 4 \\ \hline\hline
& BR $\times10^{-3}$&
$0.80 \pm 0.19$ & $0.43 \pm 0.11$ & $0.62 \pm 0.17$ & $0.37 \pm 0.09$ \\
BSW I &  $|\hH_0|^2$ &
$0.46 \pm 0.04$ & $0.001 \pm 0.006$ & $0.43 \pm 0.05$ & $0.03 \pm 0.05$ \\
& $|\hH_-|^2$ &
$0.49 \pm 0.08$ & $0.91 \pm 0.10$ & $0.57 \pm 0.05$ & $0.97 \pm 0.06$ \\ \hline\hline
& BR $\times10^{-3}$ &
$0.80 \pm 0.17$ & $0.52 \pm 0.12$ & $0.71 \pm 0.16$ & $0.52 \pm 0.11$ \\
BSW II &  $|\hH_0|^2$ &
$0.39 \pm 0.06$ & $0.07 \pm 0.04$ & $0.36 \pm 0.06$ & $0.11 \pm 0.06$ \\
& $|\hH_-|^2$ &
$0.58 \pm 0.08$ & $0.89 \pm 0.07$ & $0.63 \pm 0.06$ & $0.87 \pm 0.10$ \\ \hline\hline
& BR $\times10^{-3}$ &
$0.79 \pm 0.19 $ & $0.43 \pm 0.11 $ & $ 0.62 \pm 0.17 $ & $ 0.36 \pm 0.09 $ \\
CDDFGN &  $|\hH_0|^2$ &
$0.46 \pm 0.04$ & $0.001 \pm 0.006$ & $0.43 \pm 0.05$ & $0.031 \pm 0.05$ \\
& $|\hH_-|^2$ &
$0.49 \pm 0.08$ & $0.91 \pm 0.10$ & $0.57 \pm 0.05$ & $0.97 \pm 0.06$ \\ \hline\hline
& BR $\times10^{-3}$ &
$0.79 \pm 0.19 $ & $0.43 \pm 0.11 $ & $ 0.62 \pm 0.17 $ & $ 0.36 \pm 0.09 $ \\
AW &  $|\hH_0|^2$ &
$0.46 \pm 0.04$ & $0.001 \pm 0.006$ & $0.43 \pm 0.05$ & $0.031 \pm 0.05$ \\
& $|\hH_-|^2$ &
$0.49 \pm 0.08$ & $0.91 \pm 0.10$ & $0.57 \pm 0.05$ & $0.97 \pm 0.06$ \\ \hline\hline
& BR $\times10^{-3}$ &
$1.10 \pm 0.25$ & $0.56 \pm 0.15$ & $0.81 \pm 0.23$ & $0.47 \pm 0.12$ \\
ISGW &  $|\hH_0|^2$ &
$0.46 \pm 0.04$ & $0.001 \pm 0.006$ & $0.44 \pm 0.05$ & $0.03 \pm 0.05$ \\
& $|\hH_-|^2$ &
$0.48 \pm 0.08$ & $0.90 \pm 0.10$ & $0.56 \pm 0.05$ & $0.97 \pm 0.05$ \\ \hline\hline
\multicolumn{2}{|l|}{Experiment \cite{ref:CDF96-160E}} &
\multicolumn{4}{|l|}{ ${\cal B}(B \rightarrow \psi(2S) K^*) = ( 0.9 \pm 0.29) \times 10^{-3}$}\\ \hline
\end{tabular}

\label{tab:B-psi2S-Kstar}
\end{table}

From Table~\ref{tab:B-psi2S-Kstar} we notice that the predictions of the branching ratio, $|\hH_0|^2$ and $|\hH_-|^2$ are almost model independent. The predictions of  solutions 1 and 3 are the closest to available experimental data. The other two solutions yield low branching ratios and close to zero longitudinal polarization.

\bbc \section{Results in Factorization Approximation}
\eec
For the sake of completeness, we present in this section the factorization approximation ($\xi_1 = \xi_{21} = \xi_{V1} =1$) prediction for branching ratio, $|\hH_0|^2$ and $|\hH_-|^2$ using the five theoretical models considered for the formfactors. These predictions are presented in Table~\ref{tab:factorization}.

From Table~\ref{tab:factorization}, we see that the factorization approximation predicts low values for the branching ratios compared to experiment. For the processes $B \rightarrow \psi K^*$ and $B_s \rightarrow \psi \phi$, with the exception of BSW I model, the factorization approximation also underestimates the longitudinal polarization.

If we scale the branching ratio by a factor of 3.5, we find that the BSW I model predictions agree, within error, with the available experimental data. This could be achieved by giving the nonfactorization parameters the values,
\begin{equation}
\xi_1 = \sqrt{3.5}, \;\; \xi_{21} = \xi_{V1} =1
\end{equation}
or
\begin{equation}
\xi_1 = \xi_2 = \xi_V = \sqrt{3.5} \; .
\end{equation}
This is what has been referred to as new factorization in \cite{ref:Cheng97}.
\begin{table}
\centering
\caption{Predictions of branching ratios, $|\hH_0|^2$ and $|\hH_-|^2$ for the processes $B \rightarrow \psi K^*$, $B_s \rightarrow \psi \phi$ and $B \rightarrow \psi(2S) K^*$ using the factorization approximation $\xi_1 = \xi_{21} = \xi_{V1} =1$. The errors in branching ratios are due to the errors in Wilson coefficients, decay constants and $B$ meson life times.}

\begin{tabular}{|l|l|ccc|} \hline
\multicolumn{2}{|c|}{} & $BR \times 10^{-3}$ & $|\hH_0|^2$ & $|\hH_-|^2$ \\ \hline\hline
& BSW I & $0.40 \pm 0.24$ & 0.57 & 0.39 \\
& BSW II & $0.33 \pm 0.20$ & 0.35 & 0.64 \\
$B \rightarrow \psi K^*$ & CDDFGN & $0.24 \pm 0.14$ & 0.37 & 0.62 \\
& AW & $0.33 \pm 0.19$ & 0.12 & 0.87 \\
& ISGW & $0.15 \pm 0.09$ & 0.06 & 0.93 \\ \cline{2-5}
& Experiment \cite{ref:CLEO97} & $1.35 \pm 0.18$ & $0.52 \pm 0.08$ &
 $0.47 \pm 0.08$ \\ \hline\hline
& BSW I & $0.27 \pm 0.16$ & 0.55 & 0.41 \\
& BSW II & $0.23 \pm 0.14$ & 0.35 & 0.64 \\
$B_s \rightarrow \psi \phi$ & CDDFGN & $0.22 \pm 0.13$ & 0.32 & 0.66 \\
& AW & $0.44 \pm 0.26$ & 0.20 & 0.79 \\
& ISGW & $0.09 \pm 0.06$ & 0.20 & 0.80 \\ \cline{2-5}
& Experiment \cite{ref:CDF95,ref:CDF96} & $0.93 \pm 0.33$ & $0.56 \pm 0.21$ & - \\ \hline\hline
& BSW I & $0.24 \pm 0.14$ & 0.49 & 0.43 \\
& BSW II & $0.23 \pm 0.13$ & 0.29 & 0.69 \\
$B \rightarrow \psi(2S) K^*$ & CDDFGN & $0.13 \pm 0.07$ & 0.35 & 0.65 \\
& AW & $0.22 \pm 0.13$ & 0.23 & 0.76 \\
& ISGW & $0.14 \pm 0.08$ & 0.22 & 0.77 \\ \cline{2-5}
& Experiment \cite{ref:CDF96-160E} & $0.90 \pm 0.29$ & - & - \\ 
\hline
\end{tabular}

\label{tab:factorization}
\end{table}

\bbc \section{Discussion}
\eec
We have shown that  $B \rightarrow \psi K$ and $\psi(2S) K$ data require nonfactorized contributions in all of the five formfactor models we have considered. The smallest amount of nonfactorized contribution is needed for the BSW II model while the AW model requires the largest. We have calculated the branching ratios for $B_s \rightarrow \psi \eta$, $\psi \eta'$, $\psi(2S) \eta$, and $\psi(2S) \eta'$ in each model as functions of $\xi_{F_1}$ and displayed the result in Fig.\ \ref{fig:C}. These branching ratios averaged over the five models are tabulated in Table \ref{tab:Bs-eta-psi} for a few values of the parameter $\xi_{F_1}$.

We have used the latest CLEO data  on ${\cal B}(B \rightarrow \psi K^*)$ and the transversity amplitudes to determine the three nonfactorization parameters $\xi_1$, $\xi_2$ and $\xi_V$ defined in (\ref{eq:xis}). There are four solutions which are expressed in terms of $\xi_1$ and the ratios $\xi_{21} = \xi_2/\xi_1$, and $\xi_{V1} = \xi_V/\xi_1$. These solutions are displayed in Fig.\ \ref{fig:E} for BSW I model and in Table \ref{tab:nofactorizatiosAV} for all the five models. We find that solutions exist for $\xi_{2,1} = 1$ and $\xi_{V,1} = 1$ only in BSW I model but then $\xi_1 \neq 1$, i.e.\ there are no solutions where $\xi_1 = \xi_2 = \xi_V =1$ which would signal factorization.

Assuming that the parameters $\xi_{21}$ and $\xi_{V1}$ determined from $B \rightarrow \psi K^*$ are the same in  $B_s \rightarrow \psi \phi$ decay, we calculated the branching ratio, longitudinal and transverse polarizations of $B_s \rightarrow \psi \phi$ decay in all five models. Present data are consistent with model predictions but for a few exceptions as is seen from Table~\ref{tab:B-psi-phi}.

The branching ratio, longitudinal and transverse polarizations of $B \rightarrow \psi(2S) K^*$ decay were also calculated assuming that the parameters $\xi_{21}$ and $\xi_{V1}$ determined from $B \rightarrow \psi K^*$ are the same in  $B \rightarrow \psi(2S) K^*$ decay. The results are shown in Table~\ref{tab:B-psi2S-Kstar}.

Finally, we find that in the factorization approximation none of the formfactor models predict correctly the branching ratios for for $B \rightarrow \psi K^*$, $B_s \rightarrow \psi \phi$ and $B \rightarrow \psi(2S) K^*$. As for the longitudinal polarizations, $|\hH_0|^2$, in $B \rightarrow \psi K^*$ and  $B_s \rightarrow \psi \phi$, only the BSW I model predicts them correctly. BSW I also predicts $|\hH_-|^2$ correctly for $B \rightarrow \psi K^*$.

\vspace{0.25cm}
{\small This research was partly supported by a grant to A.N.K. from the Natural Sciences and  Engineering Research Council of Canada}.

\pagebreak


\begin{thebibliography}{99}
\bibitem{ref:Wefollow} We follow the convention $A_0 = H_0$ of ref. \cite{ref:Dighe96} rather than $A_0 = - H_0$ of ref. \cite{ref:CLEO97}.
\end{thebibliography}
\end{document}